\documentclass[aps,prd,10pt,notitlepage,showpacs,floats]{revtex4-1}
\usepackage{amsmath,amssymb,amsthm,graphicx}

\theoremstyle{plain}

\theoremstyle{definition}
\newtheorem*{definition}{Definition}
\theoremstyle{remark}

\begin{document}

\title{Dynamics of dark energy models and centre manifolds}

\author{Christian G.~B\"ohmer}
\email{c.boehmer@ucl.ac.uk}
\author{Nyein Chan}
\email{nyein.chan@ucl.ac.uk}

\affiliation{Department of Mathematics \& Institute of Origins, University College London, Gower Street, London WC1E 6BT, United Kingdom}

\author{Ruth Lazkoz}
\email{ruth.lazkoz@ehu.es}
\affiliation{Fisika Teorikoa, Euskal Herriko Unibertsitatea, 48080 Bilbao, Spain}

\date{\today}

\begin{abstract}
We analyse dark energy models where self-interacting three-forms or phantom fields drive the accelerated expansion of the Universe. The dynamics of such models is often studied by rewriting the cosmological field equations in the form of a system of autonomous differential equations, or simply a dynamical system. Properties of these systems are usually studied via linear stability theory. In situations where this method fails, for instance due to the presence of zero eigenvalues in the Jacobian, centre manifold theory can be applied. We present a concise introduction and show explicitly how to use this theory in two concrete examples.
\end{abstract}


\maketitle

\section{Introduction}
Observational cosmology provides strong evidence that the Universe is currently undergoing an epoch of accelerated expansion (see e.g.~\cite{Dunkley:2008ie}). The driving force of this expansion is called dark energy, and it appears that the Universe's energy content is made up of 74\% of this mysterious dark energy. The second biggest energy content of the Universe is a non-baryonic matter, interacting only via its gravitational attraction, called dark matter. The existence of such a substance has long been known. It is implied by the flattened galactic rotation curves observed by Zwicky as early as 1933. Neither dark energy nor dark matter have been detected directly.

The simplest model of dark energy is the cosmological constant $\Lambda$, originally introduced by Einstein in 1917 in order to construct the Einstein static universe but later abandoned. The cosmological constant appears to be in very good agreement with observational data, however, its physical interpretation is unsatisfying when adopting a particle physics point of view. In this context it is interpreted as a measure of the vacuum energy density which leads to the well known cosmological constant problem~\cite{Weinberg:1988cp}. Another important issue is the question why the energy contents of dark energy and dark matter are of similar order of magnitude today. This requires finely tuned initial conditions from which the Universe must have started its evolution.

Some of these issues can be addressed by introducing a dynamical model for the evolution of the dark energy. For instance, one could model dark energy as an evolving scalar field with an asymptotically flat potential. If dark energy interacts with dark matter, the cosmological coincidence problem may be alleviated. Therefore, various interacting dark energy models have been proposed and investigated~\cite{Wetterich:1994bg,Amendola:1999qq,Billyard:2000bh,Zimdahl:2001ar,Farrar:2003uw,Chimento:2003iea,Olivares:2005tb,Sadjadi:2006qp,Kim:2007dp,Guo:2007zk,CalderaCabral:2008bx,Boehmer:2008av,Quercellini:2008vh,Pereira:2008at,Quartin:2008px,Chen:2008ca,He:2008tn,Valiviita:2009nu}. Some models such as quintessence have been studied in~\cite{Boehmer:2009tk}. It is logical to also consider more complicated fields, like spinors~\cite{Boehmer:2009aw}, vectors~\cite{Boehmer:2007qa} and even higher order spin fields. One such suggestion was recently studied by Koivisto and Nunes~\cite{Koivisto:2009ew,Koivisto:2009fb} who introduced three-form fields and studied their cosmology. Those fields can yield a viable cosmology where they drive inflation in the early time and become the driving forces of the accelerated expansion of the Universe at late times. This has also been investigated in~\cite{Ngampitipan:2011se}.

Besides the exact form of the field describing the dark energy, one can exploit another freedom in current cosmological models. Since both dark energy and dark matter are not understood fundamentally, there are no {\it a priori} constraints on interactions between these components. In all these scenarios it seems relevant to consider some degree of interaction (other than the gravitational one) between dark matter and dark energy. Dark matter is primarily needed in the early Universe to support the formation of structure. On the other hand, dark energy is required to drive the late time accelerated expansion of the Universe. Thus, it seems plausible to study models with an energy transfer from dark matter to dark energy as time progresses. This idea has motivated a vast amount of literature on this topic, see for instance~\cite{Copeland:2006wr} and references therein. Since there are no selection criteria leading to a specific coupling, any coupling that might be studied will necessarily be phenomenological, with some models having a good physical justification while others are motivated by mathematical simplicity.

\section{Dark energy models}
\subsection{Basics}

We work in a flat FLRW spacetime whose line element is given by
\begin{align}
  ds^{2} = -dt^{2} + a^{2}(t) (dx^2+dy^2+dz^2).
\end{align}

Neglecting radiation and baryons, the conservation equations for the background dark matter (DM) fluid and the dark energy (DE) are
\begin{align}
  \dot{\rho}_{\text{DM}} &= -3 H\rho_{\text{DM}} + Q,
  \label{ev:dm}\\
  \dot{\rho}_{\text{DE}} &= -3H(1+w_{\text{DE}})\rho_{\text{DE}} - Q,
  \label{ev:de}
\end{align}
where $Q$ is an arbitrary coupling and the subscript DE stands for a generic dark energy model to be specified. The conservation equations are subject to the Friedmann constraint
\begin{align}
  H^2 = \frac{\kappa^2}{3}(\rho_{\text{DE}}+\rho_{\text{DM}}),
  \label{fried}
\end{align}
where $\kappa^2 = 8\pi G/c^2$ is the gravitational coupling strength.

\subsection{Three-forms}

Let us now consider the case where a three-form field generates both early (inflation) and late time (dark energy) acceleration. The non-zero components of the most general three-form field in this geometry is described by~\cite{Koivisto:2009fb, Koivisto:2009ew}
\begin{align}
  A_{ijk} = a(t)^{3}\, \epsilon_{ijk}X(t),
\end{align}
where $X(t)$ is a scalar function of time and $\epsilon_{ijk}$ is the standard three-dimensional permutation symbol. The equation of motion of this field is given by a modified Klein-Gordon equation, which reads
\begin{align}
  \ddot{X} = - 3H\dot{X} - V_{,X} - 3\dot{H}X - \frac{Q}{\dot{X} + 3 H X}.
\end{align}
Following~\cite{Copeland:1997et,Copeland:2006wr}, we choose an exponential potential
\begin{align}
  V(X) = V_{0} e^{-\lambda X},
\end{align}
where $\lambda$ is a dimensionless parameter and $V_{0} > 0$. Using this potential, the field equations become
\begin{align}
  \rho_{X} &= \frac{1}{2}(\dot{X}+3HX)^{2} + V(X) - Q,\\
  p_{X} &= -\frac{1}{2}(\dot{X} + 3HX)^{2} - V(X) + V_{,X}X,\\
  H^{2} &= \frac{\kappa^2}{3}(\frac{1}{2}(\dot{X}+3 H X)^{2}+V(X)+\rho_{\rm DM}),\\
  \dot{H} &=\frac{\kappa}{2}(V_{,X}X + \rho_{\rm DM}).
\end{align}

To construct a dynamical system, we use the following dimensionless variables
\begin{align}
  x &:= \frac{1}{\sqrt{6}H}(\dot{X} + 3HX),
  \label{eqn:x1}\\
  y &:= \frac{\sqrt{V}}{\sqrt{3}H},\\
  z &:= \frac{2}{\pi}\arctan\frac{3 X}{\sqrt{6}},\\
  s &:= \frac{\rho_{\rm DM}}{\sqrt{3}H}.
  \label{eqn:s1}
\end{align}
Note that we have now deviated from~\cite{Koivisto:2009fb}. Our choice of variables has the advantage that the phase space is compact by construction and we do not have to worry about the presence of critical points at infinity. These variables are motivated by noting that the Friedmann constraint now becomes
\begin{align}
  x^2 + y^2 + s^2 = 1.
\end{align}
Note that by construction $-1 \leq z \leq 1$, and moreover $-1 \leq x \leq 1$, $0 \leq y \leq 1$ and $0 \leq s \leq 1$. Our phase space is therefore a half cylinder of height 2. 

The equation of state parameter for the three-form field is defined by $w_X = p_X / \rho_X $ and thus can be written as
\begin{align}
  w_X = -1 + \frac{V_{,X}X}{\rho_X} = -1 - \frac{1}{x^2 + y^2}\sqrt{\frac{2}{3}} y^2 \lambda  \tan\left[\frac{\pi  z}{2}\right].
\end{align}
Similarly, the total equation of state parameter becomes
\begin{align}
  w_{\rm tot} &=-x^2-\frac{1}{3} y^2 \left(3+\sqrt{6} \lambda  \tan\left[\frac{\pi  z}{2}\right]\right).
\end{align}
Recall that the condition for acceleration is $w_{\rm tot} < -1/3$. We choose the coupling $Q$ to be of the form $Q = \alpha \rho_{c} H$. This results in the following autonomous system of differential equations
\begin{align}
  x' &= \frac{3}{2} x (1-x^2-y^2) + \sqrt{\frac{3}{2}} y^2 \lambda  \left(1-x \tan\left[\frac{\pi z}{2}\right]\right) - \alpha\frac{\left(1-x^2-y^2\right)}{2 x}, 
  \label{3xp}\\ 
  y' &= \frac{3}{2} y (1-x^2-y^2)  -\sqrt{\frac{3}{2}} y \lambda  \left(x+\left(-1+y^2\right) \text{tan}\left[\frac{\pi  z}{2}\right]\right),
  \label{3yp}\\
  z' &= \frac{6}{\pi}\cos\left[\frac{\pi  z}{2}\right]^2 \left(x-\tan\left[\frac{\pi  z}{2}\right]\right).
  \label{3zp}
\end{align}
Note that the equation system is invariant under the map $y\to-y$ and thus we can constrain the analysis to the $y \geq 0$ case. This is in fact not surprising since our potential is positive definite. The number of critical points of this dynamical system depends on the coupling parameter $\alpha$. Starting with $\alpha=0$, we arrive at the results of~\cite{Koivisto:2009fb} which are 

\renewcommand{\baselinestretch}{2}
\begin{table}[!ht]
\begin{tabular}[t]{|c|c|c|c|c|c|c|}
\hline
Point & $x$ & $\,y\,$ & $z$ & eigenvalues & $w_X$ & $w_{\rm tot}$\\[1ex]
\hline\hline
$A_{+}$ & $1$ & $0$ & $\frac{1}{2}$ & $ 0,-3,-3 $&$-1$& $-1$\\[1ex]
\hline
$A_{-}$ & $-1$ & $0$ & $-\frac{1}{2}$ & $ 0,-3,-3 $&$-1$& $-1$\\[1ex]
\hline
$B$ & $0$ & $0$ & $0$ & $ \frac{3}{2},\frac{3}{2},-3 $& $-1$& $0$\\[1ex]
\hline
\end{tabular}
\caption{Critical points of the uncoupled model.}
\label{nocoup_cp}
\end{table}
\renewcommand{\baselinestretch}{1}

In the presence of a coupling $\alpha \neq 0$, the number of critical points changes. Specifically, the point $B$ splits into 2 different critical points, see Figure~\ref{Fig:plot1}. The points $A_{\pm}$ remain unchanged, however, their eigenvalues do change. This is summarised in Table~\ref{coupled}.

\renewcommand{\baselinestretch}{2}
\begin{table}[!ht]
\begin{tabular}[t]{|c|c|c|c|c|c|c|}
\hline
Point & $x$ & $\,y\,$ &$z$ & eigenvalues & $w_X$ & $w_{\rm tot}$\\[1ex]
\hline\hline
$A_{+}$ & $1$ & $0$ & $\frac{1}{2}$ & $0,-3,-3+\alpha$ & $-1$ & $-1$\\[1ex]
\hline
$A_{-}$ & $-1$ & $0$ & $-\frac{1}{2}$ & $0,-3,-3+\alpha$ & $-1$ & $-1$\\[1ex]
\hline
$B_{+}$ & $\sqrt{\frac{\alpha}{3}}$ & $0$ & $\frac{2}{\pi}\arccos\left[\sqrt{3}/\sqrt{\alpha+3}\right]$ & 
$ -3,-\alpha+3,(-\alpha+3)/2$ & $-1$ & $-\frac{\alpha}{3}$\\[1ex]
\hline
$B_{-}$ & $-\sqrt{\frac{\alpha}{3}}$ & $0$ & $-\frac{2}{\pi}\arccos\left[\sqrt{3}/\sqrt{\alpha+3}\right]$ & 
$ -3,-\alpha+3,(-\alpha+3)/2$ & $-1$ & $-\frac{\alpha}{3}$\\[1ex]
\hline
\end{tabular}
\caption{Critical points of the coupled model.}
\label{coupled}
\end{table}
\renewcommand{\baselinestretch}{1}

As the coupling strength increases to its maximally allowed value, $\alpha \rightarrow 3$, the two points $B_{\pm}$ move towards the points $A_{\pm}$. When $\alpha=3$ these points merge and the system has two critical points, each with two zero eigenvalues. Note that we do not analyse this degenerate case. 

\subsection{Phantom dark energy}

Phantom dark energy models have been of interest since their first introduction by Caldwell~\cite{Caldwell:1999ew}. Cosmological observations place the dark energy equation of state close to $-1$, however, this value could also be approached from below, a case which is not excluded by observations. In a recent paper, Leon \& Saridakis~\cite{Leon:2009dt} have considered a varying-mass model for dark matter particles in the framework of phantom cosmologies. They considered a phantom dark energy model with power-law potential interacting with dark matter. Dark energy is modelled as a scalar field with negative kinetic energy whose energy density and pressure are given by
\begin{align}
  \rho_\phi &= -\frac{1}{2}\dot{\phi}^2+V(\phi),\\
  p_\phi &= -\frac{1}{2}\dot{\phi}^2-V(\phi).
\end{align}
We denote the equation of state parameter by $w_{\phi} = p_\phi/\rho_\phi$. Note that the sign of the phantom kinetic term is opposite to that of an ordinary scalar field. The dark matter energy density is assumed to depend on the mass of the dark matter particle which in turn is assumed to depend on the field $\phi$, $\rho_{\text{DM}}= M_{\text{DM}}(\phi) n_{\text{DM}}$ where $n_{\text{DM}}$ is the number density which is determined by $\dot{n}_{\text{DM}}+3H n_{\text{DM}}=0$. The field dependence of the mass can be interpreted as a coupling between the dark energy and the dark matter. Therefore, the dark matter energy density satisfies the evolution equation~(\ref{ev:dm}) where the coupling $Q$ is determined by the dependence of the mass on the field $\phi$. This means
\begin{align}
  Q = \frac{d \ln M_{\text{DM}}(\phi)}{d \phi} \dot{\phi} \rho_{\text{DM}}.
\end{align}
Assuming that the total energy-momentum tensor is conserved, the dark energy will satisfy equation~(\ref{ev:de}). Both evolution equations will be subject to the Friedmann constraint~(\ref{fried}).

Similar to~(\ref{eqn:x1})--(\ref{eqn:s1}), we define dimensionless variables as follows
\begin{align}
  x &:=\frac{\kappa\dot{\phi}}{\sqrt{6}H},\\
  y &:= \frac{\kappa\sqrt{V(\phi)}}{\sqrt{3}H},\\
  z &:= \frac{\sqrt{6}}{\kappa\phi}.
\end{align}
It is useful to express the density parameter and the equation of state in form of these variables, which gives
\begin{align}
  \Omega_{\phi} &= \frac{\kappa^2\rho_{\phi}}{3H^2} = -x^2 + y^2,\\
  w_{\phi} &=\frac{-x^2-y^2}{-x^2+y^2},\\
  w_{\text{tot}} &= -x^2-y^2.
\end{align}

In these variables, the cosmological field equations take the form of the following dynamical system
\begin{align}
  x' &= -3x + \frac{3}{2}x(1-x^2-y^2) - \frac{\lambda y^2 z}{2} - \frac{\mu}{2}z(1+x^2-y^2),\\
  y' &= \frac{3}{2}y(1-x^2-y^2)-\frac{\lambda x y z}{2},\\
  z' &= -x z^2.
\end{align}
This system possesses two physically meaningful critical points which are non-hyperbolic since there exists at least one zero eigenvalue in each of them, see Table~\ref{tab3}.

\renewcommand{\baselinestretch}{2}
\begin{table}[!ht]
\begin{tabular}[t]{|c|c|c|c|c|c|c|}
\hline
Point & $\,x\,$ & $\,y\,$ &$\,z\,$ & eigenvalues & $w_\phi$ & $w_{\rm tot}$\\[1ex]
\hline\hline
$A$ & $0$ & $0$ & $0$ & $0,3/2,-3/2$ & undefined & $0$\\[1ex]
\hline
$B$ & $0$ & $1$ & $0$ & $0,-3,-3$ & $-1$ & $-1$\\[1ex]
\hline
\end{tabular}
\caption{Critical points of the phantom dark energy model.}
\label{tab3}
\end{table}
\renewcommand{\baselinestretch}{1}

Thus, centre manifold theory is required in order to study the nature of these critical points as remarked in~\cite{Leon:2009dt}. Note that point $A$ is always unstable since we have one positive and one negative eigenvalue.

\section{Centre manifold theory}

\subsection{Mathematical background}

When discussing the mathematical background of the centre manifold theory, we closely follow Carr~\cite{Carr:1981} and Wiggins~\cite{Wiggins:1990}. In the presence of zero eigenvalues in the stability matrix at the critical point, linear theory fails to provide information on the stability of that point. The main aim of the centre manifold is to reduce the dimensionality of the system near that point so that stability of the reduced system can be investigated. There always exists an invariant local centre manifold $W^{c}$ passing through the fixed point to which the system could be restricted to study its behaviour in the neighborhood of the fixed point. The (in)stability of the reduced system determines the (in)stability of the system at that point.

Let $x\in\mathbb{R}^c$ and $y\in\mathbb{R}^s$. An arbitrary dynamical system with zero eigenvalues in the stability matrix can always be written in the following form
\begin{align}
  \dot{x} &= Ax + f(x,y), \nonumber \\
  \dot{y} &= By + g(x,y),
  \label{dyn}
\end{align} 
where
\begin{align}
  f(0,0) = 0, \qquad Df(0,0) = 0, \nonumber \\
  g(0,0) = 0, \qquad Dg(0,0) = 0.
\end{align}
Here we assume that the critical point is located at the origin and $Df$ denotes the matrix of first derivatives of the vector valued function $f$. $A$ is a $c\times c$ matrix having eigenvalues with zero real parts and $B$ is an $s\times s$ matrix having eigenvalues with negative real parts.

Note that a dynamical system with zero eigenvalues can always be rewritten into the above form by virtue of a linear change of coordinates. We will show this construction explicitly when we discuss the three-form and phantom models below. 

\begin{definition}
We call the space
\begin{align}
  W^c(0) = \{ (x,y) \in\mathbb{R}^c \times \mathbb{R}^s | y = h(x), |x| < \delta, h(0) = 0, Dh(0) = 0 \},
\end{align}
for $\delta$ sufficiently small, the centre manifold for the system~(\ref{dyn}).
\end{definition}

One can think of this manifold as the space where the stable directions $y$ are parametrised by the unstable directions $x$. One can now consider the dynamics of the system restricted to this manifold~\cite{Carr:1981}. Since the $y=h(x)$, one is left with the reduced equation
\begin{align}
  \dot{u} = Au + f(u,h(u)),\qquad u\in\mathbb{R}^c,
  \label{dynred}
\end{align}
for sufficiently small $u$. The (in)stability properties of the reduced system~(\ref{dynred}) imply (in)stability properties of the full system~(\ref{dyn}).

Next, we need to construct this centre manifold explicitly. By differentiating the defining equation $y=h(x)$ with respect to the independent variable we get $\dot{y} = Dh(x) \dot{x}$ where we used the chain rule. Eliminating $\dot{x}$ and $\dot{y}$ via~(\ref{dyn}), one arrives and the following quasilinear partial differential equation which $h$ has to satisfy
\begin{align}
  \mathcal{N}(h(x)) = Dh(x)\left[Ax+ f\left(x,h\left(x\right)\right)\right]-Bh(x)-g(x,h(x))=0,
  \label{neqn}
\end{align}
from which $h$ can be found and inserted into~(\ref{dynred}) to study the reduced system. In general it will not be possible to solve~(\ref{neqn}) analytically. However, such knowledge is in fact not needed since we are only interested in the reduced system near the critical point. Therefore, it suffices to simply assume $h$ to be of the form $h(x) = a x^2 + b x^3 + \mathcal{O}(x^4)$ and to determine to first few non-trivial terms of the Taylor expansion of $h$. Note that this is always possible and will result in a unique solution.

\subsection{A simple example}
Let us start with a simple example~\cite{Wiggins:1990} that shows the above construction. Let $(x,y)\in\mathbb{R}^2$ and consider the system
\begin{align}
  \dot{x} &= x^2y-x^5,\\
  \dot{y} &= -y+x^2,
  \label{example}
\end{align}
whose only critical point is $(x,y) = (0,0)$ with eigenvalues $0$ and $-1$. 

Therefore, in this example one has
\begin{alignat}{2}
  A &= 0, &\qquad B &= -1,\\
  f(x,y) &= x^2y-x^5, &\qquad g(x,y) &= x^2.
\end{alignat}
Assuming that $h(x)$ is of the form
\begin{align}
  h(x) = a x^2 + b x^3 + \mathcal{O}(x^4),
\end{align}
we find that 
\begin{align}
  \mathcal{N}(h(x)) &= (2ax+3bx^2+\mathcal{O}(x^4))(x^2(a x^2 + b x^3 + \mathcal{O}(x^4))-x^5)+(a x^2 + b x^3 + \mathcal{O}(x^4))-x^2 \\
  &= (a-1)x^2 + bx^3 + \mathcal{O}(x^4) = 0.
\end{align}
Comparing coefficients, we find $a=1$ and $b=0$ and thus the centre manifold is given by $h(x) = x^2 + \mathcal{O}(x^4)$.

Now, the evolution equation restricted to the centre manifold~(\ref{dynred}) is given by
\begin{align}
  \dot{x} = x^4 +\mathcal{O}(x^5),
\end{align}
whose dynamics are such that for $x$ sufficiently small, $x=0$ is unstable.

We would also like to note that it is not allowed to simply approximate the centre manifold by $y=0$. Putting this into the equations~(\ref{example}), we would arrive at the reduced equation $\dot{x} = -x^5$ which would indicate stability. More details can be found in~\cite{Carr:1981,Wiggins:1990}

\section{Applications to dark energy models}

\subsection{Three-forms}

Applications of centre manifold theory to cosmological models have been studied previously, see e.g.~\cite{Rendall:2001it,Wainwright:1997}. We now apply this to the system~(\ref{3xp})--(\ref{3zp}) for the critical point $A_{+}$. In order to do so, we must transform this system into the form of~(\ref{dyn}). Firstly, we introduce new coordinates $X=x-1$, $Y=y$ and $Z=z-1/2$ which move the point $(1,0,1/2)$ to the origin of the phase space. By computing the matrix of eigenvectors of the stability matrix of the system in $X,Y,Z$, we introduce another set of new coordinates
\begin{align}
  \begin{pmatrix} u \\v \\w \end{pmatrix} = 
  \begin{pmatrix}
    0 & 1 & 0\\
    -3/(\pi\alpha) & 0 & 1\\
    3/(\pi\alpha) & 0 & 0
  \end{pmatrix}
  \begin{pmatrix} X \\Y \\Z \end{pmatrix}.
\end{align}
In these coordinates, our system of equations is now in the correct form
\begin{align}
  \begin{pmatrix} \dot{u} \\ \dot{v} \\ \dot{w} \end{pmatrix} = 
  \begin{pmatrix}
    0 & 0 & 0 \\
    0 & -3 & 0 \\
    0 & 0 & -3 + \alpha 
  \end{pmatrix}
  \begin{pmatrix} u \\ v \\ w \end{pmatrix}
  + 
  \begin{pmatrix} \text{non} \\\text{linear} \\\text{terms} 
  \end{pmatrix}.
\end{align}
Comparing this with the general form~(\ref{dyn}), we firstly note that $x=u$ is a scalar function while $y=(v,w)$ is a two-component vector. Accordingly, we find
\begin{align}
  A &= 0, \qquad
  B = \begin{pmatrix}
    -3 & 0 \\
    0 & -3 + \alpha 
  \end{pmatrix},\\
  f &= -\frac{1}{6}\pi\alpha (6+\sqrt{6}\lambda)u w-
  \frac{3}{2}u^3 -\frac{1}{6} \pi^2 \alpha^2 u w^2 -  
  \sqrt{\frac{3}{2}} u \lambda  \left(1+\left(u^2-1\right) \tan \left[\frac{1}{2} \pi  (\frac{1}{2}+v+w)\right]\right), \\
  g &= \begin{pmatrix}
    (\alpha + 2 a_2 \pi\alpha-3)u^2 + 3 a_3 u^3 + \mathcal{O}(u^4) \\
    (3 + 2 b_2 \pi\alpha)u^2 + b_3 (3-\alpha) u^3 + \mathcal{O}(u^4)
  \end{pmatrix}.
\end{align}
Note that knowledge of $g$ up to order $ \mathcal{O}(u^4)$ is sufficient to construct the centre manifold up to the desired order. The complete expressions for the components of $g$ can be found in Appendix~\ref{appA}.

\begin{figure}[!ht]
\centering
\includegraphics[width=0.48\textwidth]{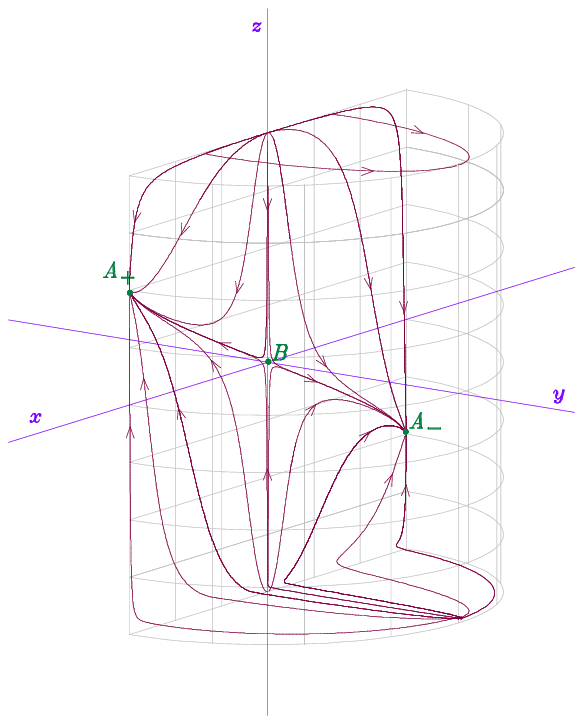}
\includegraphics[width=0.48\textwidth]{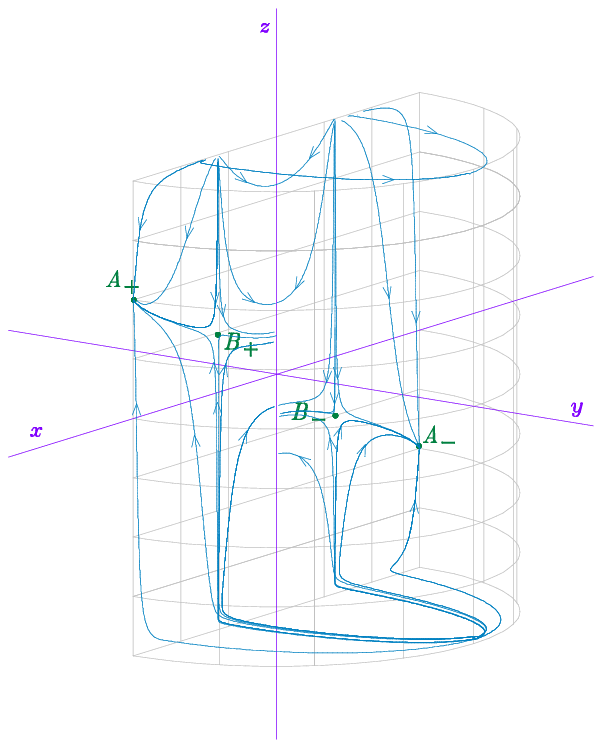}
\caption{These figures show the phase space of the three-form dark energy models without (left) and with dark sector couplings (right).}
\label{Fig:plot1}
\end{figure}

The centre manifold can now be assumed to be of the form 
\begin{align}
  h = \begin{pmatrix}
    a_2 u^2 + a_3 u^3 + \mathcal{O}(u^4)\\
    b_2 u^2 + b_3 u^3 + \mathcal{O}(u^4)
  \end{pmatrix}.
\end{align}
It has to satisfy equation~(\ref{neqn}), which explicitly reads
\begin{align}
  \mathcal{N} = \frac{1}{2\pi\alpha}\begin{pmatrix}
    3(\alpha + 2 a_2 \pi\alpha-3)u^2 + 3 a_3 u^3 + \mathcal{O}(u^4),\\
    (3-\alpha)(3 + 2 b_2 \pi\alpha)u^2 + b_3 (3-\alpha) u^3 + \mathcal{O}(u^4),
  \end{pmatrix} = 0.
\end{align}
Solving for the four constants $a_2,a_3,b_2$ and $b_3$, we obtain
\begin{alignat}{2}
  a_2 &= \frac{3-\alpha}{2\pi\alpha}, &\qquad a_3 &= 0,\\
  b_2 &= \frac{-3}{2\pi\alpha}, &\qquad b_3 &=0.
\end{alignat}
We can now study the dynamics of the reduced equation~(\ref{dynred}), which becomes
\begin{align}
  \dot{u} = -\left(\frac{3}{2}+\sqrt{\frac{3}{2}}\alpha\right) u^3 + \mathcal{O}(u^4).
\end{align}
Therefore, we find that the point $A_{+}$ is stable according to centre manifold theory. When we repeat this calculation for $A_{-}$ we find the opposite results, namely, this point is unstable. Figure~\ref{Fig:plot1} shows the phase spaces of the coupled and the uncoupled models. Note that this information is not obvious from the trajectories in the phase space diagrams since one is tempted to conclude that both $A_{+}$ and $A_{-}$ are stable.  

\subsection{Phantom dark energy}

We extend our work to a scenario of varying-mass dark matter particles in the framework of phantom cosmology, and our results obtained by performing centre manifold analysis are consistent with~\cite{Leon:2009dt}.

We firstly look at model $2$ of~\cite{Leon:2009dt}. The autonomous system of differential equations is given by
\begin{align}
  x' &= -3x + \frac{3}{2}x(1-x^2-y^2)-\frac{\lambda y^2 z}{2}-\frac{\mu}{2}z(1+x^2-y^2),\\
  y' &= \frac{3}{2}y(1-x^2-y^2)-\frac{\lambda xyz}{2},\\
  z' &= -xz^2.
\end{align}
The critical points are $(0,0,0)$ and $(0,1,0)$, which are both non-hyperbolic. However, the former has a positive and a negative eigenvalue, from which it can be deduced the point is unstable. As for the second point, there are two negative eigenvalues and a zero eigenvalue. Therefore, linear theory fails to provide information about the stability of that point. It is required to perform centre manifold analysis in order to study the nature of the stability of the second point. We first transform the coordinates into a new system $X = x, Y = y - 1$ and $Z = z$, such that the critical point is at the origin. Next, we introduce yet another new coordinate system
\begin{align}
  \begin{pmatrix} u \\v \\w \end{pmatrix} = 
  \begin{pmatrix}
    0 & 1 & 0\\
    1 & 0 & \frac{\lambda}{6}\\
    0 & 0 & 1
  \end{pmatrix}
  \begin{pmatrix} X \\Y \\Z \end{pmatrix}.
\end{align}
In these new coordinates the equations are transformed into
\begin{align}
  \begin{pmatrix} \dot{u} \\ \dot{v} \\ \dot{w} \end{pmatrix} = 
  \begin{pmatrix}
    0 & 0 & 0 \\
    0 & -3 & 0 \\
    0 & 0 & -3 
  \end{pmatrix}
  \begin{pmatrix} u \\ v \\ w \end{pmatrix}
  + 
  \begin{pmatrix} \text{non} \\\text{linear} \\\text{terms} 
  \end{pmatrix}.
\end{align}
As before, we compare this with the general form~(\ref{dyn}) and deduce that $x = u$ is a scalar function while $y = (v,w)$ is a two-vector component vector. Thus, we obtain
\begin{align}
  A &= 0, \qquad
  B = \begin{pmatrix}
    -3 & 0 \\
    0 & -3 
  \end{pmatrix},\\
  f &= -\frac{1}{6} u^2 (6 v-u \lambda ),\\
  g &= \begin{pmatrix} g_1 \\ g_2 \end{pmatrix},
  \label{phantom:g}
\end{align}
where we give the explicit expressions of the components of $g$ in Appendix~\ref{appB}.

The centre manifold can now be assumed to be of the form 
\begin{align}
  h = \begin{pmatrix}
    a_2 u^2 + a_3 u^3 + \mathcal{O}(u^4)\\
    b_2 u^2 + b_3 u^3 + \mathcal{O}(u^4)
  \end{pmatrix}.
\end{align}
Equation~(\ref{neqn}) becomes
\begin{align}
  \mathcal{N} = \begin{pmatrix}
    3 a_2 u^2+\frac{1}{144} u^3 \left(432 a_3+72 b_2 (\lambda -2 \mu )-\lambda ^2 (4+\lambda -2 \mu )\right)+ \mathcal{O}(u^4),\\
    3 b_3 u^3+u^2 \left(3 b_2-\frac{\lambda ^2}{24}\right) + \mathcal{O}(u^4),
  \end{pmatrix} = 0,
\end{align}
where the $a_2, a_3, b_2$ and $b_3$ are computed as
\begin{alignat}{2}
  a_2 &= 0, &\qquad a_3 &=  \frac{\lambda ^2}{108},\\
  b_2 &= \frac{\lambda ^2}{72}, &\qquad b_3 &=0.
\end{alignat}

The dynamics of the system restricted to the centre manifold is given by
\begin{equation}
  \dot{u} = \frac{\lambda}{6}u^3 + \mathcal{O}(u^4).
\end{equation}
Thus it is clear from this equation that the point is stable if $\lambda <0$ and unstable if $\lambda>0$. This result is consistent with \cite{Leon:2009dt} where the same result was obtained by using the method of normal forms.

\section{Conclusions}

One of the main aims of this paper was to give a concise introduction into centre manifold theory, which is useful for studying dynamical systems with zero eigenvalues in the stability matrix. While this method has recently been applied to some models, a simple introduction with applications to cosmology does not seem to appear in the literature. We provide such an introduction and point out the necessity of using centre manifold theory.

We then apply this method to two dark energy models, one based on three-forms and the other one based on phantom dark energy. For the phantom dark energy model we confirm previous results and show their derivation explicitly. For the three-form model we can conclude that only one of two potentially stable points is in fact stable, a result not accessible from linear stability theory. Moreover, the trajectories of the phase space seem to suggest stability of the unstable point, emphasising that critical points with zero eigenvalues need to be investigated carefully. 

In order to study physically relevant dynamical systems derived from cosmological field equations, it is important that we do not restrict our models based on simplicity and the ability to use linear stability theory. We should aim towards constructing the most physically motivated models and study them irrespective of the mathematical challenges involved. There exists a huge amount of mathematical literature on dynamical systems, much of which still awaits to be applied to physics.

\begin{acknowledgments} 
NC would like to thank Ben Willcocks for useful discussions on centre manifold theory and suggestions on the manuscript. RL is supported by the Spanish Ministry of Science and Innovation and by the Basque Governments through research projects FIS2010-15492 and GIU06/37 respectively.
\end{acknowledgments}

\appendix

\section{Explicit formula of the three-form model}
\label{appA}

The two components of $g$ in the three-form model are given by
\begin{multline}
g_1 = \frac{1}{6} \left(\frac{9}{\pi }+18(v+w) +3 w \alpha(1 +3 \pi  w) +\pi ^2 w^3 \alpha ^2+\frac{27}{\pi  (3+\pi  w \alpha )}-\frac{6}{\pi}(3 \cos[\pi  (v+w)]- (3+\pi  w \alpha ) \sin[\pi  (v+w)])\right)\\
+\frac{1}{6} u^2 \left(9 w-\frac{27}{\pi  (3+\pi  w \alpha )}+\frac{3}{\pi\alpha}\left( 3\left(3-\sqrt{6} \lambda \right)+ \sqrt{6}(3+\pi  w \alpha ) \lambda  \tan\left[\frac{1}{2} \pi  \left(\frac{1}{2}+v+w\right)\right]\right)\right),
\end{multline}
\begin{multline}
g_2 = \frac{3 u^2 \left((3-6 \pi  w) \alpha -\pi ^2 w^2 \alpha ^2 - 9\right)}{2 \pi  \alpha  (3+\pi  w \alpha )}\\
- \frac{1}{\pi\alpha}\sqrt{\frac{3}{2}} u^2 \lambda  \left(3-(3+\pi  w \alpha ) \tan\left[\frac{1}{2} \pi  \left(\frac{1}{2}+v+w\right)\right]\right) \\
- \frac{\pi  w^2 \alpha  \left(27+3 (1+4 \pi  w) \alpha +\pi ^2 w^2 \alpha ^2\right)}{6 (3+\pi  w \alpha )}.
\end{multline}

\section{Explicit formula of the phantom model}
\label{appB}

The two components of $g$ are given by
\begin{multline}
  g_1 = -\frac{3}{2} v \left(v^2+w (2+w)\right)+\frac{1}{4} u \left(w (2+w) (2\mu-\lambda)+v^2 (3 \lambda -2 \mu )\right)\\
  -\frac{1}{24} u^2 v \lambda  (4+3 \lambda -4 \mu )+\frac{1}{144} u^3 \lambda ^2 (4+\lambda -2 \mu ),
\end{multline}
and 
\begin{align}
  g_2 = -\frac{3}{2} \left(v^2 (1+w)+w^2 (3+w)\right)+\frac{1}{24} u^2 (1+w) \lambda^2.
\end{align}

\end{document}